# Multiple Silicon Dangling-Bond Charge qubits for quantum computing: A Hilbert-Space Analysis of the Hamiltonian

Zahra Shaterzadeh-Yazdi[1]

[1] School of Engineering Science, College of Engineering, University of Tehran, Tehran, Iran

E-mail: zahra.shaterzadeh@ut.ac.ir

**Abstract**

Silicon-based dangling-bond charge qubit is one of the auspicious models for universal fault-tolerant solid-state quantum computing. In universal quantum computing, it is crucial to evaluate and characterize the computational Hilbert space and reduce the complexity and size of the computational space. Here, we recognize this problem to understand the complexity and characteristics of the Hilbert space in our dangling-bond qubit model. The size of the desired Hilbert space can prominently be reduced by considering assumptions regarding the qubit loss. Moreover, the dimension of the desired subsets in the space shrinks by a factor of two due to the spin preservation property. Finally, the required classical memory for storage of the qubit information, Hamiltonian and Hilbert space is analysed when the number of qubits grows.

Keywords: Dangling-Bond Pair (DBP⁻), Charge Qubit, Hamiltonian, Hilbert Space Analysis, Representation theory

## 1. Introduction

Quantum computing (QC) has been one of the most trending fields in the last decade. Computational power that QC brings with itself has motivated researchers to look for a large scalable universal quantum computing system [1]. Implementing high-quality and well-characterized qubit, as the most crucial building block of QC, is the main challenge to overcome. Different hardware platforms and testbeds have been introduced throughout the years. Realizing a semiconductor solid-state qubit, especially in silicon (Si), is one of the primary states of interest. This is due to the maturity of Si technology and the compatibility of future QC chips with current computing chips.

The frontrunner silicon-based qubit candidates are electron spin [2–5] and charge qubits [6, 7]. Charge qubits are realized using superconducting cooper-pairs, but they are hard to implement and suffer from serious decoherence effects [8, 9]. Electron-spin qubits, which are based on the spin of charge carriers in semiconductors, have less decoherence challenge, but they are hard to control and measure. An approach to read out spin qubits is to initially convert them into charge qubits [10]. Quantum state of a charge qubit can be encoded in a pair of coupled quantum dots (QDs), possessing an extra electron. The electron located on the left (and right) QD can be considered as the logical state $|0\rangle$ (and $|1\rangle$). Superposition of the logical states that is generally of the form $\alpha|0\rangle + \beta|1\rangle$, occurs by coherent tunneling between the two QDs.

One type of QD charge qubit, which was realized in recent years, is the silicon-based coupled dangling-bond pair, also known as DBP⁻ [11]. Coupling between dangling bonds was first observed experimentally in 2009 [12]. Later, coupled dangling-bond pairs were proposed as charge qubit and their characterization showed that they are promising qubit families for solid-state quantum computing [11, 14]. A Si-based dangling bond (DB) is generated by selectively removing a hydrogen atom from a hydrogen-terminated silicon surface, by means of a scanning tunneling microscope. A dangling-bond charge qubit is comprised of two closely-spaced DBs sharing an excess electron that tunnels back and forth between the two DBs, and consequently coupling them with each other.

Dangling bond is known as a truly-atomic size quantum dot, due to its structural and electronic features [12]. Reducing the size of quantum dots to atomic scale helps in detracting the separation distance between them, thus increasing significantly the amount of tunneling rate, while the decoherence rate scales weakly. It is shown that for coupled DB pairs, the ratio of tunneling rate and decoherence rate is enormously large and overpass the fault-tolerance threshold [11]. Therefore, the DB charge-qubit model is a promising candidate for fault-tolerant quantum computation. Moreover, such configuration establishes DB quantum dots on H-terminated surface instead of complicated heterostructures QDs. Further research on the qubit dynamic model [11], characterizing the coherence rate [13], and characterization of coupling strength [14] in $DBP^-$ qubits have already been investigated.

As the number of DB qubits grows for the purpose of universal quantum computing, it is crucial to investigate and characterize the computational Hilbert space associated with the system and to reduce the complexity and size of the computational space. Here we recognize and tackle this problem to understand the complexity and characteristics of the desired and undesired Hilbert space in our $DBP^-$ qubit model. By taking into account the assumptions regarding the qubit loss, the size of the desired Hilbert space can prominently be reduced. Moreover, the dimensionality of the desired subsets in the space shrinks by a factor of two due to the spin preservation property. Finally, the required classical memory for storage of the qubit information is analyzed when the number of qubits grows.

 Our paper is organized as follows. Section 2 describes the basics and physical description of the $DBP^-$ qubit. Additionally, the quantum dynamics and Hamiltonian of the qubit are represented using the extended Hubbard model. Finally, the origins of qubit losses are introduced and discussed. Section 3, analyzes the size and complexity of the Hilbert space and utilizes the assumptions regarding the qubit loss and spin preservation to reduce the size of the desired Hilbert space. The required memory for storage of the qubit information is calculated for various assumptions. Finally, Section 4 concludes the paper and discusses future works.

## 2. Dangling-Bond Pair Charge Qubits

The silicon surface of interest is the hydrogen-terminated Si(100)–2×1. This surface is known as an attractive platform for nanoscale patterning. Each silicon atom of the surface is capped by a hydrogen atom and it shares one bond with another surface silicon; it also contributes two bonds with the silicon atoms located within the Si bulk. A dangling bond is created when a hydrogen atom is abolished from a Si atom of the surface by a scanning probe [15–17]. The energy state of the resulted DB is located within the Si crystal bandgap and possesses one confined electron. Accordingly, DB is a truly atomic scale quantum dot [12], as its energy state is separated from the crystal conduction and valence bands; hence it displays localized features. Depending on the doping type and the temperature, a DB can obtain an extra electron or release its own electron to the bulk and, respectively, become negatively charged ($DB^-$) or positively charged ($DB^+$). The crystal under study in this work is a highly-concentrated phosphorous-doped (n-type) crystal. Thus, every DB sustains an excess electron and becomes negatively charged.

It was experimentally shown that if two negatively-charged DBs are created within a distance of 16Å from each other, they exhibit coupling behavior and show coherent quantum dynamics, realizing them to serve as a good candidate for Si-based charge qubit [11, 12]. In fact, when a $DB^-$ is located close to another $DB^-$ (within 16Å), the enormous coulomb repulsion from the excess electron in one of the $DB^-$s restricts the other DB from hosting its extra charge. As depicted in Fig. 1, DBs arrangement on the Si surface will lead to a dangling-bond pair ($DBP^-$) with one shared electron tunneling between them. The tunneling rate is proportional to the separation distance in $DBP^-$ and the geometry of the Si surface. For the minimum and maximum separation 3.84Å and 16Å of $DBP^-$, the ab-initio calculated tunneling rates are $\simeq 467 THz$ and $\simeq 0.1 THz$, respectively [13, 14]. Since the energy levels of a $DBP^-$ are localized within the Si-crystal bandgap, the $DBP^-$ can be modeled as two potential wells, with the local position of the shared electron being represented by the right or the left state [11]. Equivalently, with a proper basis transformation, the two lowest energy levels of the extra electron can also represent symmetric and anti-symmetric states. Moreover, the coupling of the DBP- with its surrounding environment is negligible [12], thus, the electron spin is assumed to be preserved. Using this assumption, the Hilbert space dimension can be folded by a factor of two which will be discussed and used in Section 3.

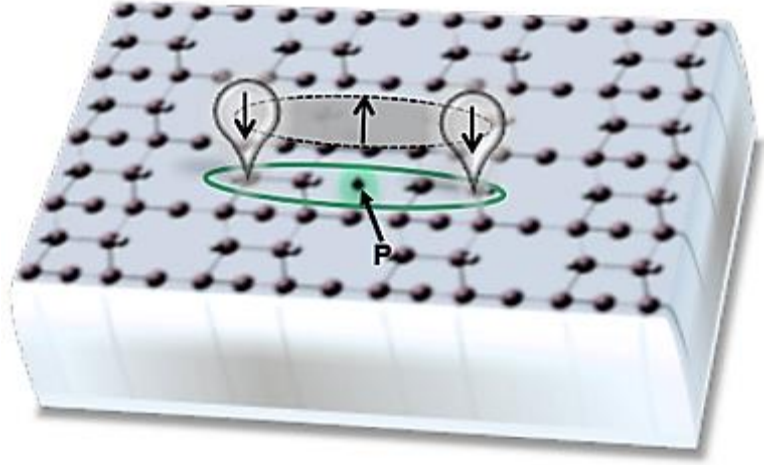

Figure 1. A DBP⁻ on a hydrogen-terminated Si (100)-2x1 surface. Bubbles represent dangling bonds that are coupled with each other by an extra charge provided by a Phosphorus dopant (P). Each DB possesses an electron; arrows show the spin up/down of the electron.

For one DB, a Hamiltonian applied to the Hilbert space spanning zero, one, or two electrons per lattice site on the silicon surface, can express the charge qubit dynamics. The Hamiltonian model includes the tunneling of electrons, Coulomb repulsion between electrons (on-site and inter-DB), on-site energy ($E_{os}$) and potential diffrences ($\hat{V}$) across the surface [11]. For an arbitrary number of qubits (dangling bond at position i), the system's Hamiltonian on a two-dimensional surface can be described as:

$$\hat{H} = \sum_{i,\sigma} E_{os} \hat{n}_{i,\sigma} - \sum_{\langle i,j \rangle, \sigma} T_{ij}(\hat{c}^\dagger_{i,\sigma}\hat{c}_{j,\sigma} + \hat{c}^\dagger_{j,\sigma}\hat{c}_{i,\sigma}) + \sum_i U_i\, \hat{n}_{i,\uparrow}\hat{n}_{i,\downarrow} + \sum_{\langle i,j \rangle, \sigma, \sigma'} W_{i\sigma j\sigma'}\hat{n}_{i,\sigma}\hat{n}_{j,\sigma'} + \hat{V} \qquad (1)$$

Where $\langle i,j \rangle$ denote the nearest-neighbor interactions at the lattice sites $i$ and $j$, and $\{\sigma, \sigma'\} \in \{\uparrow, \downarrow\}$ are the up and down electron spins. Furthermore, $\hat{c}^\dagger_{i,\sigma}$, $\hat{c}_{i,\sigma}$ and $\hat{n}_{i,\sigma}$ are creation, annihilation, and number operators, respectively, that act on electrons with spin $\sigma$ at lattice-site $i$. The tunneling between positions $i$ and $j$ are given by $T_{ij}$. The on-site Coulomb interaction $U_i$ indicates the energy cost of placing two electrons with opposite spin to the same position $i$. The inter-site Coulomb effect denoted by $W_{i\sigma j\sigma'}$ is the cost of placing another spin $\sigma'$ electron at site $j$ while simultaneously placing a spin $\sigma$ electron at site $i$. Finally, $\hat{V} = \frac{1}{2}\sum_{i<j,\sigma} V_{ij}(\hat{n}_{i,\sigma} - \hat{n}_{j,\sigma})$ is the potential difference across the surface of the Silicon.

In DBP⁻s, qubit loss occurs due to different mechanisms. One of the crucial causes of qubit loss is tunneling of excess electrons to other neighboring dangling bonds. This loss can be reduced by placing all of the DBP⁻s in appropriate positions from each other. As an example, for 3.84Å of distance between DBs in a DBP⁻, the tunneling rate and Coulomb repulsion between electrons are 0.308eV and 0.583eV, respectively. Nevertheless, when two DBP⁻s are placed about 17.92Å from each other, the tunneling rate drops to $0.128 \times 10^{-4}$eV which is negligible. Thus, one can tune the tunneling rate and qubit loss by placing the DBP⁻s at appropriate positions.

Additionally, qubit loss also happens when the excess charge of the DBP⁻ transfers to higher energy levels that are located in the conduction band of the Si crystal. When two coupled DBs are at the shortest-possible distance from each other (i.e. 3.84Å), the DBP⁻ is in its highest energy level. For this case, our calculation shows that the probability of the excess electron to go to conduction band (qubit loss) is around $1.453 \times 10^{-8}$ which is negligible. Moreover, in some cases, the excess charge, provided by the dopant, for instance by phosphorous atom (P), recombines with the dopant atom and leads to loss of qubits in DBP⁻s. The electron-donor recombination rates also depend on the DB-pair separation. Our calculations show that the probability of losing an electron to a P atom (qubit loss), in the worst case, is around $1.11 \times 10^{-5}$ in this case, which is insignificant.

In conclusion, in all three possible cases of qubit loss, placing the DBs at proper positions ensures that qubit loss is not significant and can be neglected. Such an assumption will benefit the dimensionality reduction of the Hilbert space and could provide a simplified desirable Hilbert space to perform quantum computation.

## 3. Hilbert Space Analysis

For utilizing the full potential of the DBP⁻-model of the qubits and performing quantum computing using this model, a complete analysis of the Hilbert space is necessary. The size of the Hilbert space for N qubit is $\binom{4N}{3N}$. The total Hilbert space consists of two desired and undesired part and can be represented by H $=\mathbb{H}_d + \mathbb{H}_{und}$, where *d* and *und* stand for desired and undesired part of the Hilbert space, respectively. The undesired Hilbert space corresponds to the cases where qubit loss happens, i.e. when a DBP⁻ loses its excess electron.

The desired Hilbert space has a dimension of $4^N = 2^N \times 2^N$, where one of the $2^N$ represents the number of subspaces (blocks) in the desired Hilbert space and the other $2^N$ corresponds to the dimension of each subspace. The Hilbert space dimension can equivalently be written as $4^N = N(\oplus\, 2^N) = 2^N \oplus \ldots \oplus 2^N$.

Table 1. Desirable, undesirable, and total Hilbert space sizes for various number of qubits in the DBP⁻ model.

| For 1 Qubit | $\binom{4}{3} = 4 = 2^1 \times 2^1 + 0$ |
|---|---|
| For 2 Qubits | $\binom{8}{6} = 28 = 16 + 12 = (2^2 \times 2^2) + 12$ |
| For 3 Qubits | $\binom{12}{9} = 220 = 64 + 156 = (2^3 \times 2^3) + 156$ |
| … | … |
| For N Qubits | $\binom{4N}{3N} = \frac{4N!}{3N!\,N!} = (2^N \times 2^N) + \left(\binom{4N}{3N} - (2^N \times 2^N)\right)$ |

(Desired States + Undesired States)

A few examples for demonstrating such property are given in Table 1. It can be seen that for one DBP⁻ qubit, the Hilbert space is 4 dimensional composed of 2 subspaces, each being 2 dimensions, without any undesirable subspace. For 2 DBP⁻ qubits, the total Hilbert space has 28 dimensions, where the desired part has 16 dimensions and the undesired part has 12 dimensions. For 3 qubits, the dimension of Hilbert space increases to 220, where the desired Hilbert space is 64 dimensional and the rest, i.e. 156, corresponds to the dimension of the undesired Hilbert space.

In order to find the Hamiltonian of N-qubit system using the representation theory, one needs to apply the fermionic annihilation and creation operators, $\hat{C}_{NLR\uparrow\downarrow}$ and $\hat{C}^\dagger_{NLR\uparrow\downarrow}$, to the Hilbert space of $\binom{4N}{3N-1}$ dimension, where $N \in \{1, 2, \ldots\}$, which corresponds to N number of DBP⁻s, each possessing 3 electrons. We note that it is not possible to go from 3N+1 subspace to 3N subspace by applying the annihilation operator.

The question arises is that how many number of basis states of the $\binom{4N}{3N-1}$ Hilbert space is required in order to find the desired Hamiltonian of N qubits. In order to find the answer, we need to expand the equation which corresponds to the total number of required basis states consisting of desired and undesired blocks of the N-qubit Hamiltonian, which is given by

$$\binom{4N}{3N-1} = \frac{4N!}{(3N-1)!(N+1)!} = 3N2^{2N-1}. \qquad (2)$$

Therefore, the total number of basis states required to expand the Hilbert space of N number of DBP⁻s is given by $(3N-1)!\,(N+1)!$, where $3N2^{2N-1}$ of them is the basis set required to construct the desired block of the N number of DBP⁻ (qubit) Hamiltonian, and the rest of states construct the undesired block of the Hamiltonian.

For the whole N-qubit Hilbert space, the number operator is $\hat{n} = \hat{c}^\dagger \hat{c}$, where the dimension of $\hat{n}$ is $2^{2N} \times 2^{2N}$ and the dimension of $\hat{c}$ is $3N2^{2N-1} \times 2^{2N}$. Operator $\hat{c}^\dagger$ is the adjoint of the annihilation operator, and its Hilbert space dimension is $2^{2N} \times 3N2^{2N-1}$. For each block subspace N qubit Hilbert space, considering spin preservation in each block, the dimension of $\hat{n}$ is $2^N \times 2^N$ and the dimension of $\hat{c}$ is $5N2^{N-1} \times 2^N$.

Table 2. Summary of the results regarding the dimension of the required states for different cases and assumptions, as the number of qubits/electrons grows.

| # of qubits / # of electrons | $\mathbb{H} = \mathbb{H}_d + \mathbb{H}_{und}$ | $\mathbb{H}_{Spin}$ | $\mathbb{H}_{Shared}$ | $\mathbb{H}_{Unshared}$ | $(\hat{n} = \hat{c}^\dagger \hat{c})$ | $(\hat{n} = \hat{c}^\dagger \hat{c})_{spin}$ |
|---|---|---|---|---|---|---|
| 1 Qubit 3 Electrons | $\binom{4}{2}=6+0$ | 5 | 4 | 1 | $4 \times 4 = (4 \times 6)(6 \times 4)$ | $2 \times 2 = (2 \times 5)(5 \times 2)$ |
| 2 Qubit 6 Electrons | $\binom{8}{5}= 56=48\oplus 8$ | 20 | 16 | 4 | $16 \times 16 = (16 \times 48)(48 \times 16)$ | $4 \times 4 = (4 \times 20)(20 \times 4)$ |
| 3 Qubit 9 Electrons | $\binom{12}{8}= 495=288\oplus 207$ | 60 | 48 | 12 | $64 \times 64 = (64 \times 288)(288 \times 64)$ | $8 \times 8 = (8 \times 60)(60 \times 8)$ |
| … | … | … | … | … | … | … |
| N Qubit 3N Electrons | $\binom{4N}{3N-1} = 3N2^{2N-1} \oplus \mathbb{H}_{und}$ | $5N2^{N-1}$ | $4N2^{N-1}$ | $N2^{N-1}$ | $2^{2N} \times 2^{2N} = (2^{2N} \times 3N2^{2N-1})(3N2^{2N-1} \times 2^{2N})$ | $2^N \times 2^N = (2^N \times 5N2^{N-1})(5N2^{N-1} \times 2^N)$ |

Table 2 summarizes all of the simplified dimensions of the Hilbert space and subspace blocks as the number of DBP⁻ qubits grows. The second column indicates the total number of states (desired + undesired) required to construct the Hamiltonian of N qubit system using the representation theory, i.e. using $\hat{n}$, $\hat{c}^\dagger$, $and$ $\hat{c}$. For instance, for the case of one qubit, the dimension is 6, all of which are needed to construct the desired part of the Hamiltonian. For two qubit case, the dimension is 56, where 48 are for reconstructing the desired subspace of the Hamiltonian and 8 is to construct the undesired subspace of the Hamiltonian. The third column represents the total number of states required to construct each block subspace (constrained to spin preservation) and is stated as $\mathbb{H}_{Spin}$. The fourth and fifth columns (stated as $\mathbb{H}_{Shared}$ and $\mathbb{H}_{Unshared}$, respectively) represent the total number of states shared between each subspace and N other subspaces and the number of unshared states in each subspace, respectively. Finally, the last two columns represent the dimension of the number operator, the annihilation operator and the creation operator without considering spin preservation (6th column) and for constructing each block subspace considering spin preservation (7th column).

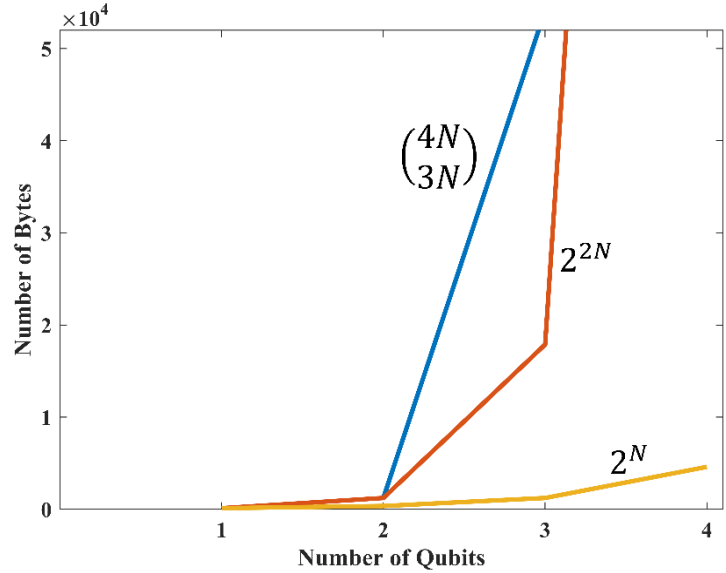

Figure 2. The classical memory required to store the qubit information for different qubit spaces. The blue line indicates the whole Hilbert space (without taking into account the assumptions). The orange line represents the required storage when qubit loss is negligible (with spin) and the yellow line is when spin preservation and qubit loss are both taken into account and the space is even more simplified. As evident, the required classical resources are significantly decreased because the size of the desirable Hilbert space is reduced.

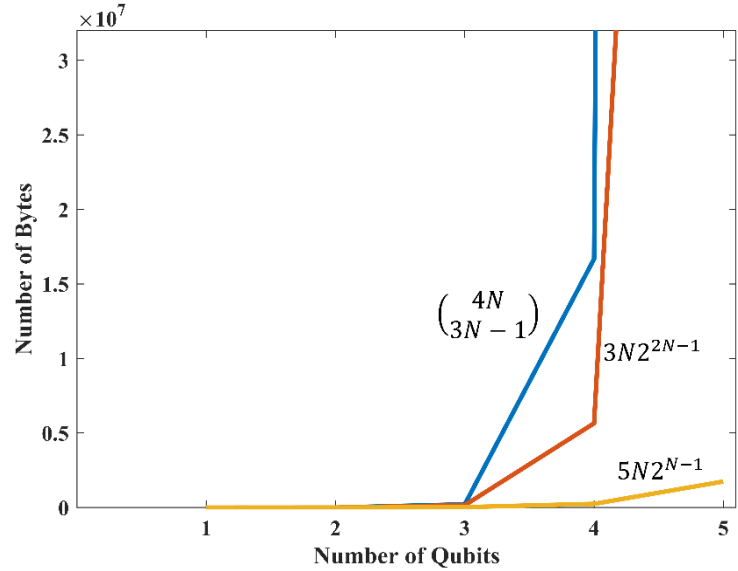

Figure 3. Blue line is the classical memory (number of bytes) required to store the total number of states (desired + undesired) which construct the Hamiltonian of N qubit system. The orange line is the size of the desired states for constructing the Hilbert space and the yellow line is the total number of states needed to construct each block subspace (constrained to spin preservation). As evident, the required classical resources is significantly decreased because the size of the necessary states to construct the Hamiltonian is reduced.

In order to analyze and perform quantum computing via DBP⁻ model of qubits, it is also crucial to determine the size of the required storage memory to restore the qubit information. Thus, it is valuable to calculate the amount of storage, i.e. the number of Bytes needed to keep the qubit information. We show that, considering our qubit loss and spin preservation assumptions, the simplified Hilbert space will consume significantly reduced amount of the classical storage and the minimal resources are required. Figure 2 illustrates a comparison of the required classical memory to restore the size of the Hilbert space for different assumptions. As evident, the number of Bytes required to restore the information of the Hilbert space plunges dramatically as the number of qubits grows. Hence, the analysis of the Hilbert space and attain a simplified desirable space will also use minimum classical resources.

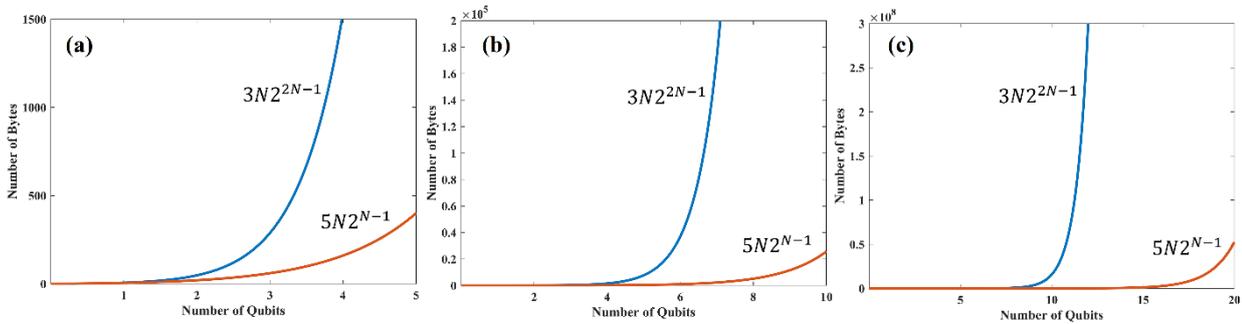

Figure 4. The classical memory required to store the desired states for constructing the Hilbert space (blue line) and the total number of states needed to construct each block subspace constrained to spin preservation (orange line). (a) for up to 5 qubits, (b) for up to 10 qubits and (c) for up to 20 qubits. As evident, the classical resources are reduced exponentially as the number of qubits grows.

Moreover, Fig. 3 indicates the required classical memory to restore the number of states needed for construction of the Hamiltonian of an N qubit system. It can be seen that the required memory is significantly reduced for the case of no qubit loss and spin preservation. Working with a simplified smaller Hilbert space will lead to huge optimization in classical resources. Same argument holds for Fig. 4 and Fig. 5, which depict the required memory to restore the desired states of the Hilbert space and the number operator, respectively.

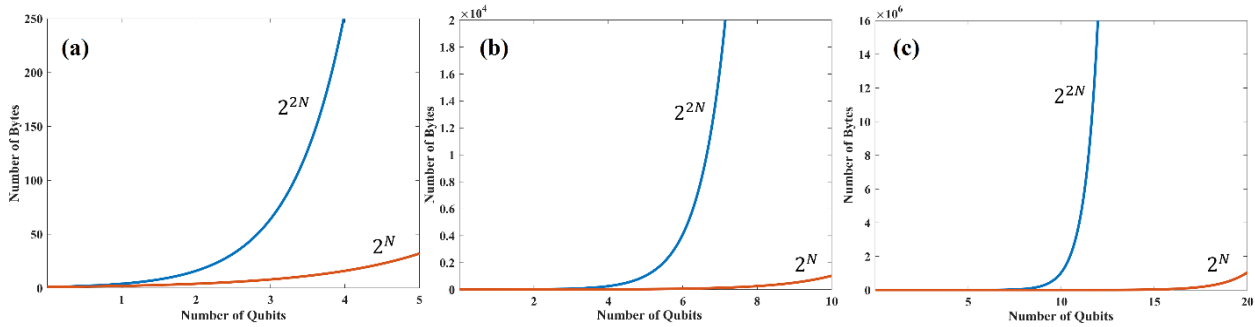

Figure 5. The classical memory required to store the number operator without spin preservation (blue line) and with spin preservation (orange line). (a) for up to 5 qubits, (b) for up to 10 qubits and (c) for up to 20 qubits. As evident, the classical resources are reduced exponentially as the number of qubits grows.

## 4. Conclusion

In this research work, we characterized the computational Hilbert space associated with dangling-bond pair (DBP⁻) model of charge qubit, to be used as a unit of information in solid state quantum computing. The DBP⁻ qubit model has high tunneling rate, negligible qubit loss and spin preservation during the computational process. Thus, by analyzing the Hilbert space associated with such a qubit, we managed to utilize the qubit loss and spin preservation assumptions to simplify the Hilbert space and only use the desirable part of the Hilbert space. At the desirable part of the Hilbert space, qubit loss is negligible and spin is preserved, which makes the DBP⁻ model of the qubit a prominent candidate for quantum computing purposes. Moreover, the classical resource analysis was given to evaluate the classical memory required to restore the information of the qubits, Hilbert space, and operators that construct the Hamiltonian of the system.